\documentclass[aps,10pt,floats,floatfix,showpacs,amssymb,prd,twocolumn,superscriptaddress,nofootinbib]{revtex4-2}

\usepackage{graphicx}% Include figure files
\usepackage{dcolumn}% Align table columns on decimal point
\usepackage{bm}% bold math
\usepackage{amsmath}
\usepackage[colorlinks=true,linkcolor=blue,citecolor=blue,urlcolor=blue]{hyperref}
\usepackage{dsfont}
\usepackage{listings}
\usepackage{xcolor}

\newcommand\comment[1]{}

\begin{document}

%\preprint{APS/123-QED}

\title{Evolution of self-gravitating spherical dark-matter halos with and without new physics}

\author{Marc Kamionkowski}
\email{kamion@jhu.edu}
\affiliation{William H.\ Miller III Department of Physics \& Astronomy, Johns Hopkins University, 3400 N.\ Charles St., Baltimore, MD 21218, USA}

\author{Kris R.\ Sigurdson}
\email{krs@phas.ubc.ca}
\affiliation{Department of Physics and Astronomy, University of British Columbia, Vancouver, BC V6T 1Z1, Canada}

\setcounter{footnote}{0}
\def\thefootnote{\arabic{footnote}}

\begin{abstract}
We present an efficient numerical algorithm for evolving self-gravitating systems of dark-matter particles that leverages the assumption of spherical symmetry to reduce the nominally six-dimensional phase space to three dimensions.  It can be used to quickly determine numerically the evolution of an initially static stable self-consistent self-gravitating system if there is some additional or new physics.  We illustrate here with four examples:  (1) the effects of the growth of a supermassive black hole at the center; (2) the effects of stripping of the outer layers of the halo (a toy model for the effects of tidal stripping of galaxies); (3) the response of a self-gravitating system to dark matter that decays to a slightly less massive state; and (4) the effects of a slow change to Newton's constant. The approach can be extended to study dark matter with elastic and inelastic self-interactions and to study the process of virialization in spherical collapse.  We describe some aspects of a code {\tt NSphere} that implements this approach which we are making available.
\end{abstract}

\maketitle
%\tableofcontents

\section{Introduction}

Often, detailed numerical calculations and simulations are required to attempt to match the level of detail of observations and measurements of complicated astrophysical systems.

Sometimes, though, a simplified model can be of value, particularly when exploring new physics, for which the space of models of interest may be huge.  It is often advisable to understand, for example, the evolution of spherical systems before attempting to model realistic systems that may not be precisely spherical.  

With this spirit, we present here a new approach to studying the evolution of self-gravitating spherical systems of collisionless particles.\footnote{See, e.g., Ref.~\cite{Binney:2008} for a clear pedagogical overview.}  Such a system, for example, a galactic dark-matter halo \cite{Freeman:1970mx,Rubin:1970zza,Ostriker:1974lna,1974Natur.250..309E,Rubin:1980zd}, is comprised of particles that have orbits that move in a gravitational potential generated self-consistently by the mass distribution implied by that collection of orbits.  A spherical halo of a given density profile can be constructed from a distribution function (DF) that depends only on the energies and angular momenta of the particles that comprise the system.  The relation between the DF and the density profile is given by Eddington's formula.  Such a system can be shown under fairly general circumstances to be stable \cite{Doremus:1971zz}.

If, however, there is some new physics, or if the system is somehow perturbed, the delicate balance between the particle orbits and the gravitational potential they move in can be disrupted.  A change in orbits results in the mass distribution and thus the potential and consequently affects all the other orbits, which then again change the density distribution and potential, repeating the cycle.  Although qualitative descriptions of the evolution can be provided, the evolution is not amenable to any precise analytic description and has typically been studied with $N$-body codes.

Here, we describe a simple and efficient technique to numerically evolve self-gravitating spherical systems.  The idea is that the six-dimensional phase-space distribution can, for spherical systems, be collapsed to three dimensions: the particle's distance from the origin, its velocity, and the angle the velocity makes with respect to the radial direction.  Moreover, the force on any given particle is always radially directed and determined entirely by the enclosed mass.  The computational time of the calculation thus scales as $N \ln N$ with the number $N$ of particles, rather than $N^2$ as in a traditional $N$-body simulation.  Fluctuations are smaller given that the $N$ particles are distributed in a 3d, rather than 6d, phase space, and there is no concern about relaxation due to strong two-particle interactions.  It is easily coded and allows those without supercomputing resources or GPUs (or patience) to study gravitational dynamics in spherical systems.

This approach is not new---it was described originally by H\'enon in 1964 \cite{Henon:1964}(but differs from the H\'enon's later Monte Carlo technique \cite{Henon:1971a,Henon:1971b} employed in many star-cluster codes \cite{Joshi:1999vf, Giersz:2011em, Rodriguez:2018pss}---but it has gone under the radar among the community of people studying galactic halos.  The approach is also inspired by $N$-1-body codes for the nonlinear evolution of non-cold relics \cite{Ringwald:2004np,Brandbyge:2010ge,Mertsch:2019qjv,Worku:2024kwv}.

Below we detail the approach, then illustrate it with several examples.  The first illustrates a model: dark matter decays to a slightly less massive state, thus giving the decay product a small velocity kick \cite{Peter:2010au,Peter:2010xe,Wang:2014ina,Cheng:2015dga,Chen:2020udm}.  A second example is the evolution of a halo if there is time evolution in Newton's constant (or, equivalently, time evolution in the dark-matter particle mass).  We then obtain the change in the density profile in the event of a slow growth of a SMBH at the galactic center.  The last example is the response of a self-gravitating system to the removal of its outer layers, a toy model for the response of a halo to tidal stripping.  The result shows a relaxation to a system with a density profile that drops at large radii $r$ as $r^{-10}$.

We also describe a code, {\tt NSphere}\footnote{\url{https://github.com/kris-sigurdson/NSphere}}, that uses this algorithm.  It is a fairly straightforward implementation in C, but there are a few features that are worth calling out.  It is easily imaginable that with more effort further computational efficiencies in this code can be incorporated.

This paper is organized as follows:  Section \ref{sec:initialconditions} describes how the initial conditions for a self-consistent self-gravitating system of particles are determined for a given density profile.  Section \ref{sec:eoms} then describes how the system is evolved forward in time.  Section \ref{sec:ddm} illustrates by studying the evolution of a galactic halo composed of dark matter that decays to a slightly massive state and a relativistic particle (that escapes the system).  Section \ref{sec:variableG} explores the evolution of a halo in which Newton's constant is slowly changing with time; this also describes roughly the evolution of a halo in which dark matter decays or annihilates to particles that escape the system.  Section \ref{sec:stripping} provides perhaps the most novel and interesting example, a halo in which the outer 40\% of the mass has been suddenly stripped.  The remaining mass is seen to slosh back and forth and ultimately relax to a density profile that approaches a constant at small radii and falls off as $r^{-10}$ at large radii.  The final example, in Section \ref{sec:smbh} shows the growth of a cusp around a slowly growing supermassive black hole at the center of the halo.  Section \ref{sec:conclusions} summarizes and describes some other possible directions to be explored with this new approach.  Two Appendices describe computational algorithms we have developed for {\tt NSphere} to integrate low-angular-momentum orbits and to parallelize the sorting of particles by radii that must occur at each time step.

\section{Initial conditions}
\label{sec:initialconditions}

\subsection{The self-consistent distribution function}

Starting with a spherically symmetric self-gravitating system of density $\rho(r)$  that is $\rho(r)=0$ at radii $r>R$ and has total mass $M_h=M(R) \equiv 4\pi\int_0^R\, dr\,r^2\,\rho(r)$.  Following Ch.\ 4 of Ref.~\cite{Binney:2008} we define a relative gravitational potential $\Psi(r) = -\Phi(r) + \Phi_0$, with $\Phi(r)$ the usual gravitational potential, so that the Poisson equation is $\nabla^2\Psi(r)= -4\pi G\rho(r)$ and take the boundary condition $\Psi(r) \to 0$ as $r\to \infty$.  The potential has a radial gradient $\partial \Psi(r)/\partial r = (4\pi G/r^2) M(r)$

The halo is composed of particles with velocity $v$ and kinetic energy per unit mass $(1/2)v^2$ and potential energy per unit mass at position $r$ given by $\Phi(r)$.  We then define a relative energy ${\cal E} = \Psi(r)-(1/2)v^2$.

The mass density is $\rho(r)$ of dark-matter particles is obtained by integrating the distribution function (DF) over the velocity.  If the halo is spherically symmetric, the DF is most generally a function of the particle energy ${\cal E}$ and its angular momentum, but the simplest assumption is that it is a function $f({\cal E})$ simply of the energy.  If so, the DF is obtained from the density profile through Eddington's formula,
\begin{equation}
     f({\cal E}) = \frac{1}{m_\chi\sqrt{8}\pi^2} \frac{d}{d{\cal E}}\int_0^{\cal E}\, \frac{d\Psi}{\sqrt{{\cal E}-\Psi}} \frac{d\rho}{d\Psi},
\end{equation}
which follows from the relation,
\begin{eqnarray}
   \rho(r) &=& m_\chi \int\, d^3 v\, f\left[\Psi(r) -\frac12 v^2 \right] \nonumber \\
   &=& 4\pi m_\chi \int_0^{\Psi(r)} \, d{\cal E} \,f({\cal E}) \sqrt{ 2[\Psi(r) -{\cal E}]},
\end{eqnarray}
and $m_\chi$ is the mass of the dark-matter particle. If the DF does not depend on the angular momentum, then the velocity distribution is isotropic everywhere in the halo.

\subsection{Generating the realization of the distribution}

We now construct a population of $N$ particles with radii $r_i$ and velocities $v_i$ (with $i=1,.\ldots,N$) that follows the desired distribution.  To do so, we first generate for each particle $i$ a radius $r_i$ chosen from a cumulative probability distribution $(1/N)(dN/dr) = M(r)/M$, with $M(r)=4\pi \int_0^r\, dr\, r^2\, \rho(r)$.  We then invert this distribution and obtain a random radius drawn from this distribution by picking a random number between 0 and 1.  Once the radius $r_i$ is picked, we must draw a velocity from a distribution,
\begin{equation}
     \left.\frac{1}{N}\frac{dN}{dr}\right|_r \propto v^2 f \left[ \Psi(r) - v^2/2\right].
\end{equation}
This distribution has a different functional form for each value of $r$, and evaluating the cumulative distribution function for each $r$ would be too computationally expensive.  Instead, we choose a random $v_i$ between 0 and $\sqrt{2\Psi(r)}$ and then another random number $x$ in the range $0<x<1$.  If \begin{equation}
    x<\frac{v^2 f\left[\Psi(r)-v^2/2 \right]}{ \left (v^2 f\left[\Psi(r)-v^2/2 \right]\right)_{\rm max}},
\end{equation}
then the velocity is assigned to be $v_i$.  If not, we keep choosing random pairs of $v_i$ and $x$ until this condition is satisfied.  The maxima $ \left (v^2 f\left[\Psi(r)-v^2/2 \right]\right)_{\rm max}$ can be precomputed as a function of $r$.

% The other option is a Metropolis-Hastings algorithm.  Although this is less computationally intensive, the samples it draws will have correlations that may decrease the stability of the self-gravitating system.

Once the set of $(r_i,v_i)$ is chosen, we assign each particle velocity a random polar angle $-\pi/2<\theta_i<\pi/2$ by choosing $\mu_i=\cos\theta_i$ from a uniform distribution between $-1$ and 1.

\begin{figure}
\includegraphics[width=\columnwidth]{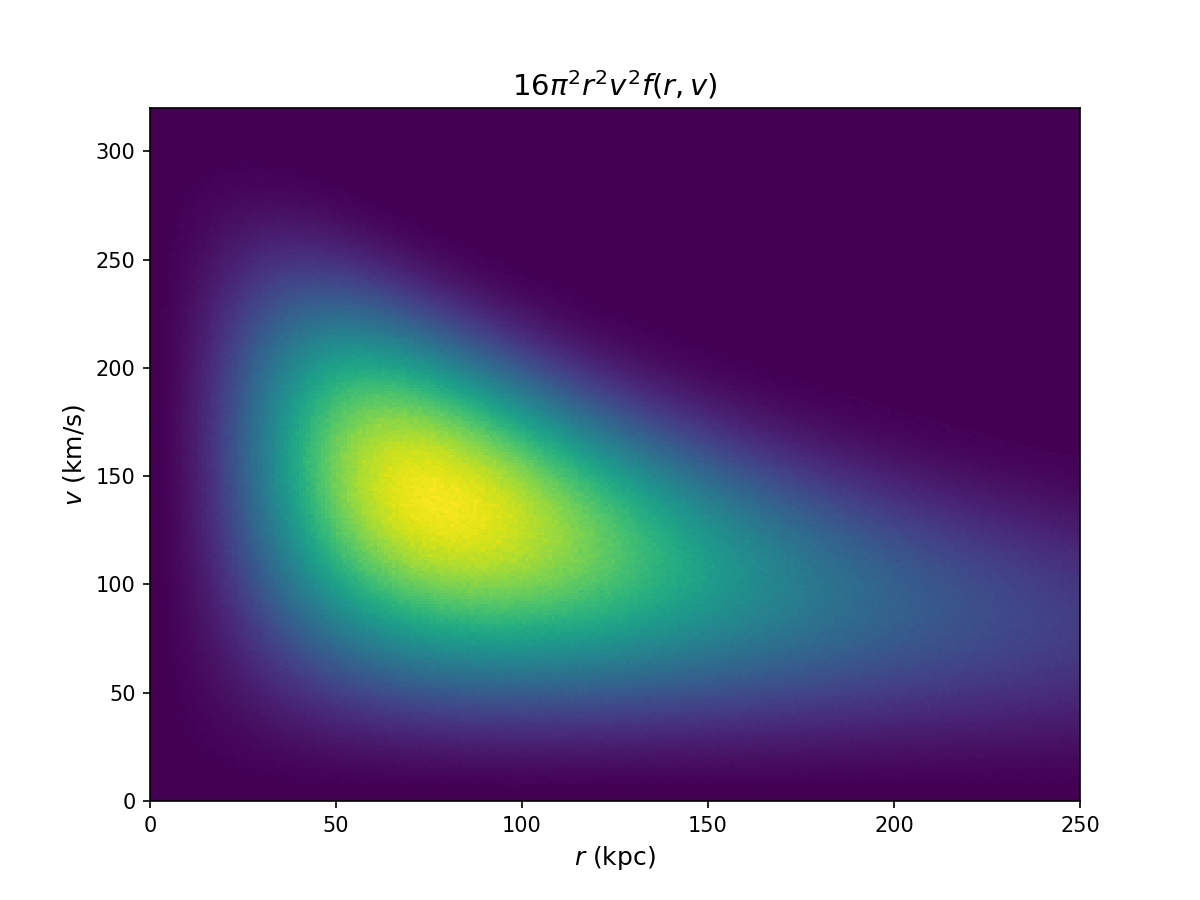}
   \caption{The initial phase space distribution $16 \pi^2 r^2 v^2 f(r,v)$ versus radius $r$ and velocity $v$, generated by sampling the theoretical DF (Section~\ref{sec:initialconditions}) using 200,000,000 particles.}
\label{fig:DF}
\end{figure}

In the numerical results presented here, we have taken the halo density profile to be $\rho(r) \propto (1+r^2/r_c)^{-3}$, where $r_c$ is a core radius, but the calculation is easily modified to work with any density profile.  Fig.~\ref{fig:DF} shows an example of an initial distribution $r^2 v^2 f(r,v)$ of particles in a simulation.

\section{Evolution with time}
\label{sec:eoms}

With spherical symmetry, each particle has a distance $r_i$ from the origin governed by the equation of motion,
\begin{equation}
     \ddot r = \frac{\partial \Psi}{\partial r} + \frac{\ell^2}{r^3},
\end{equation}
where $\ell$ is the angular momentum per unit mass.  We evolve the system forward in time with the particle position evolved at each small timestep by the equation of motion, and the potential evaluated from the mass distribution.  We first re-write the second-order equation as two first-order equations (for $\dot r$ and $r$).  The simplest approach is then to step forward using a leapfrog integration.  Here, the radius $r_i$ and radial velocity $v_{r,i}$ at discrete time steps are labeled by an integer $i$.  In each iteration, we then first advance $r'=r_i+v_{r,i} \Delta t/2$, where $\Delta t$ is the time step.  We then re-order all the particles so that they are in order of increasing $r$.  The mass $M(r)$ enclosed at any given $r$ is then the particle mass times the number of particles with smaller radius.  The velocity is then advanced $v_{r,i+1} = v_r -\left[G M(r')/r'^2 + \ell^2/r'^3\right] \Delta t$.  The radius is then advanced another half step, $r_{i+1}=r' + v_{r,i+1} \Delta t/2$.  With $N$ particles in the simulation, the ultimate precision will be $O(N^{-1/2})$, and so, roughly speaking, we need to have (given that the leapfrog algorithm is second order) the time step to be $\Delta t \lesssim N^{-1/4} t_{\rm dyn}$.

For particles in low-angular-momentum orbits, the equations of motion can briefly become stiff when the particle is near pericenter.  The leapfrog integration, with fixed stepsize, then breaks down.  We have thus implemented in the code {\tt NSphere} an adaptive stepper, described in Appendix \ref{sec:stepper}, to deal with these cases.  The sorting of the $N$ particles at each time step is a $\sim N\ln N$ calculation that is typically not easily parallelized.  However, with small step sizes, the particles are very nearly ordered in each step which then allows some parallelization, as described in Appendix \ref{sec:sort}, thus affording some speedup of the code.

\begin{figure}
\includegraphics[width=\columnwidth]{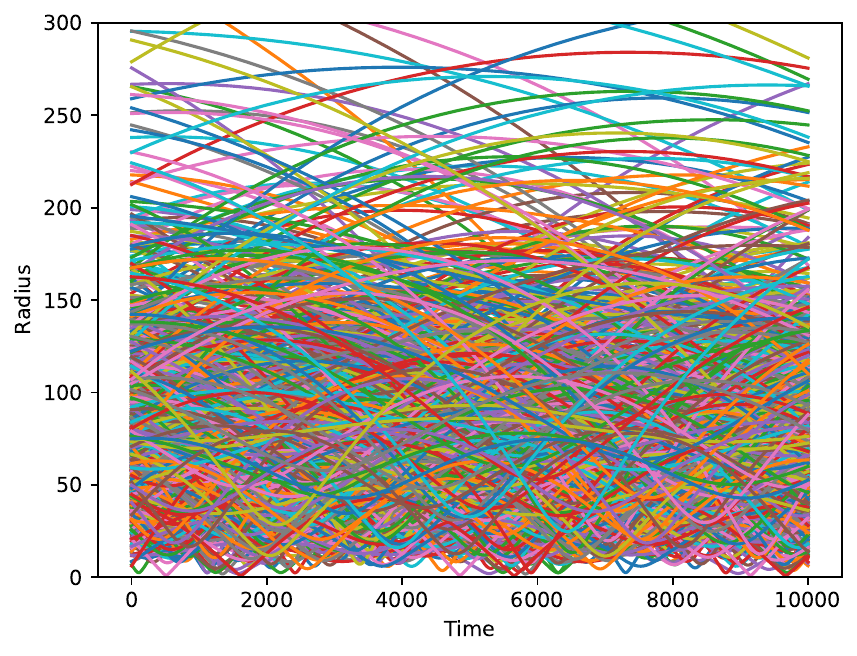}
     \caption{Here we show an example of the ``orbits'' in the simulation.  More precisely, we plot the trajectories which are here described by a radius as a function of time, for 1000 randomly chosen particles in a simulation of 100,000 particles. }
\label{fig:orbits}
\end{figure}

To illustrate, we show in Fig.~\ref{fig:orbits} the trajectories of a number of particles in a given simulation.  

\section{Decaying dark matter}
\label{sec:ddm}

\begin{figure}
\includegraphics[width=\columnwidth]{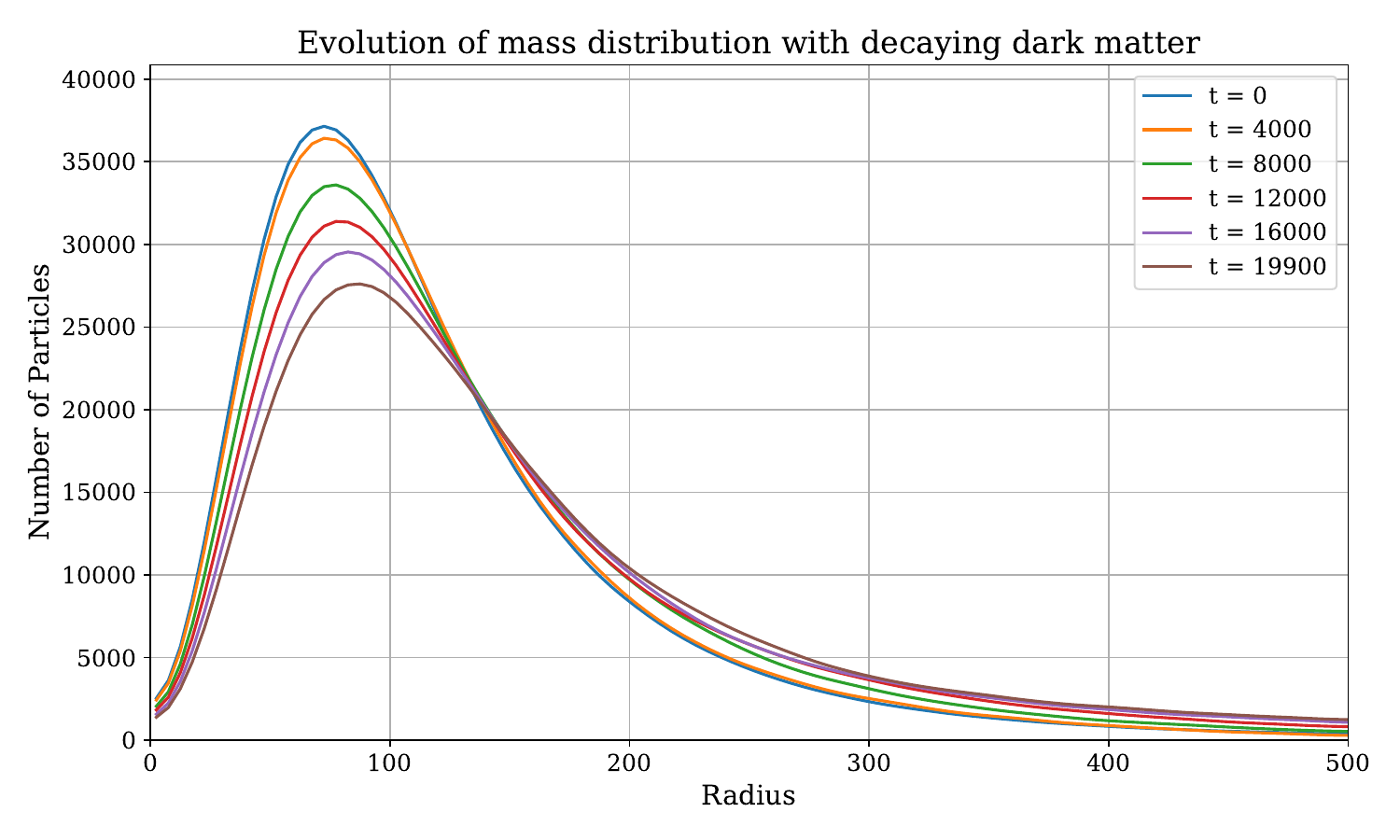}
     \caption{The time evolution of the mass distribution ($\propto r^2\, \rho(r)$ (in dimensionless units) of an initially self-consistent self-gravitating system with decaying dark matter.  More precisely, in each time step of width $\Delta t$, each particle has a chance $\Gamma \Delta t$ to receive a randomly oriented velocity change of magnitude $\Delta v=100$~km~sec$^{-1}$.  Here, the lifetime $\Gamma^{-1}$ has been taken to be twice the integration time of 20 dynamical times, with 1000 time steps per dynamical time.  There were $10^6$ particles in this simulation.}
\label{fig:decayingdm}
\end{figure}

We first illustrate the utility of this calculation by studying the evolution of a self-gravitating dark-matter halo composed of decaying dark matter.  Consider a population of dark-matter particles of mass $m_\chi$ that are initially in an excited state with a lifetime $\Gamma$ for decay to a state with mass $m_\chi - \Delta m$ with $\Delta m \ll m_\chi$ and another (noninteracting) massless particle.  The massless particle quickly escapes the system and carries negligible energy.  Thus, the dark-matter particle gets a velocity kick $\Delta {\bf v}$ of magnitude $\Delta v \simeq (\Delta m/m_\chi)c$.  The DDM parameter space can then be taken to be $\Gamma$ and $\Delta v$.  

To approximate this scenario we simply give each particle a probability $\Gamma \Delta t$ to receive a randomly oriented velocity change $\Delta {\bf v}$ of some fixed magnitude.  The cosmologically relevant parameter space spans values of $\Gamma^{-1}$ that are long compared to the age of the Universe  (but not too much longer) to values that are shorter than the age of the Universe, but only by an order of magnitude or so (if $\Gamma^{-1}$ is too short, the particles are all decayed before the halo forms).  The interesting values of $\Delta v$ are in the ballpark of the halo velocity dispersion, which ranges from tens of km/sec for smaller subhalos to $\sim1000$ km/sec for galaxy clusters.  Interesting halo masses should span the range of $\sim10^6\, M_\odot$ for the earliest collapsed objects to $\sim10^{15}\, M_\odot$ for galaxy clusters.

Here we illustrate with the example of a halo of $10^{12}\,M_\odot$ with a scale radius $r_c=100$ kpc, and thus particle velocities on order 200 km~sec$^{-1}$.  We integrate for 20 dynamical times and take the decay lifetime to be twice the integration time (so that roughly half the particles have received a kick over the integration time).  We simulated $10^6$ particles.   Results for the evolution of the density profile are shown in Fig.~\ref{fig:decayingdm}.   We leave a more detailed study of the functional evolution of $\rho(r)$ to future work, but qualitatively, the results make sense:  The heating due to particle decays causes the halo to expand (and also ejects some particles).

\begin{figure}
\includegraphics[width=\columnwidth]{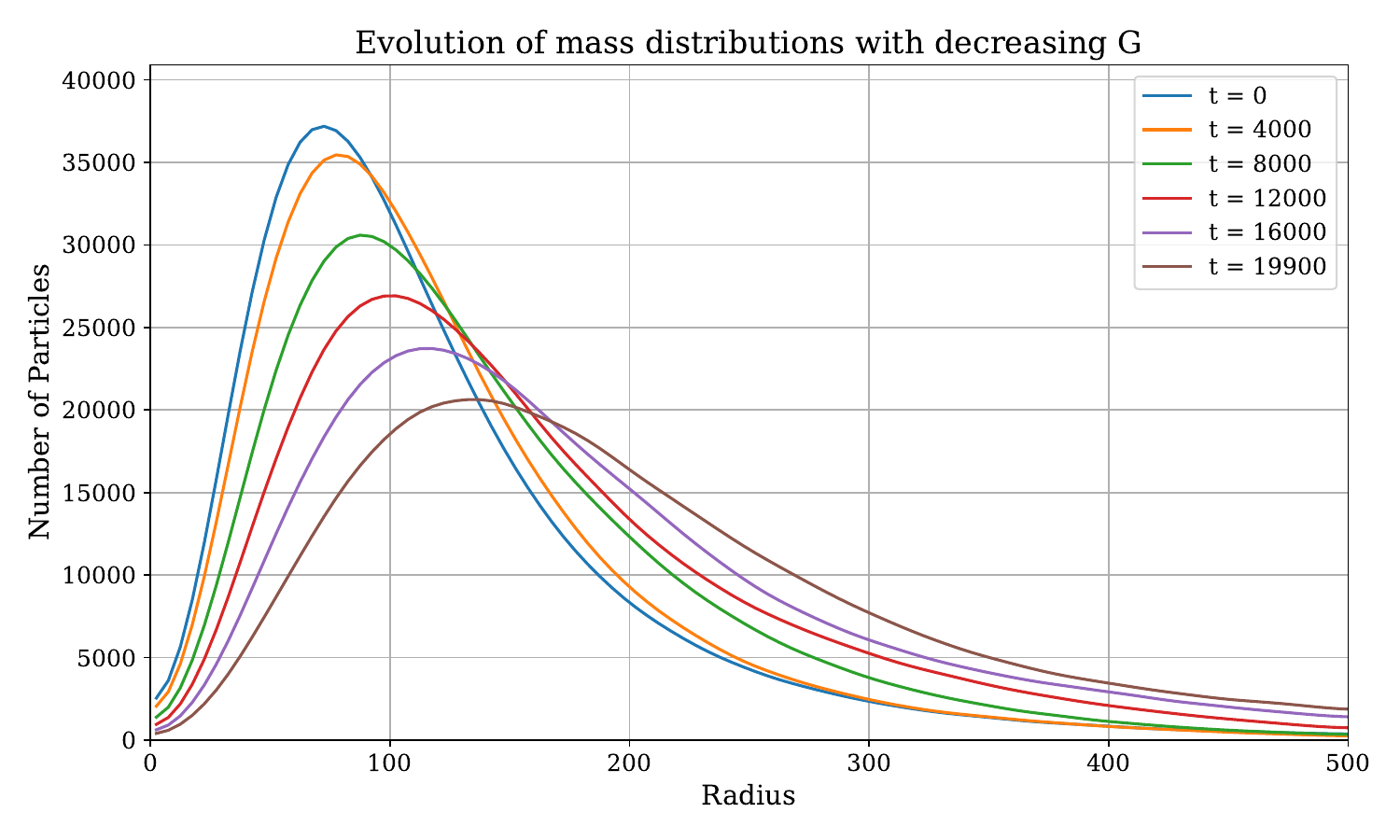}
     \caption{The time evolution of the mass distribution ($\propto r^2\, \rho(r)$ (in dimensionless units) of an initially self-consistent self-gravitating system with a time evolving Newton's constant $G$.  Here, $G$ decreases linearly by a factor of 2 over the integration time.}
\label{fig:evolvingG}
\end{figure}

\section{Time variable Newton's constant}
\label{sec:variableG}

We now carry out the same exercise for a model in which the gravitational force constant $G$ decreases by a factor of 2 over the age of the Universe.  The analogous results are shown in Fig.~\ref{fig:evolvingG}.  The results are not too surprising: The distribution gets stretched in radius by a factor $\propto G^{-1}$.  One could also try a sudden change in $G$, where the evolution would not be adiabatic.  Rather than do so, though, we consider in the next Section another system that is suddenly thrown out of equilibrium.  The evolution seen in Fig.~\ref{fig:evolvingG} also describes roughly that for a halo in which the dark matter decays to particles that escape the system, given that the reduction in mass in this case is degenerate with the changing Newton's constant.  The only difference is that the time evolution would be exponential, rather than linear, as in Fig.~\ref{fig:evolvingG}.

\section{``Tidal stripping''}
\label{sec:stripping}

It is well known in theory of stellar structure that the response of a spherical self-gravitating fluid system (i.e., star) to the rapid removal of its outer layer depends on the entropy profile of the star \cite{Webbink:1985}.  With some entropy profiles, the remaining star expands, and with others, it contracts.  It is interesting to ask the same question about self-gravitating fluid systems of collisionless particles.  If we suddenly strip the outer layers, does the remaining system expand? or contract?  i.e., if we remove all particles at some radius $r>R_1$, where $M(R_1)$ encloses some fraction $f$ of the total mass, what happens to the remaining $1-f$ particles?

Particles in the stripped layer have relative energies in the range $0<{\cal E}<{\cal E}_{\rm strip}$. Particles at smaller radii in this relative-energy range would travel, before the stripping, to these radii, but with reduced mass, they may travel to even larger radii.  Depending on the initial profile and the amount of mass stripped, some particles may even escape the system.  The remaining particles will move in a potential that has been altered by the reduction of mass and so conceivably redistribute themselves to larger radii.  But how much larger?  And how in detail will they redistribute themselves?

\begin{figure}
\includegraphics[width=\columnwidth]{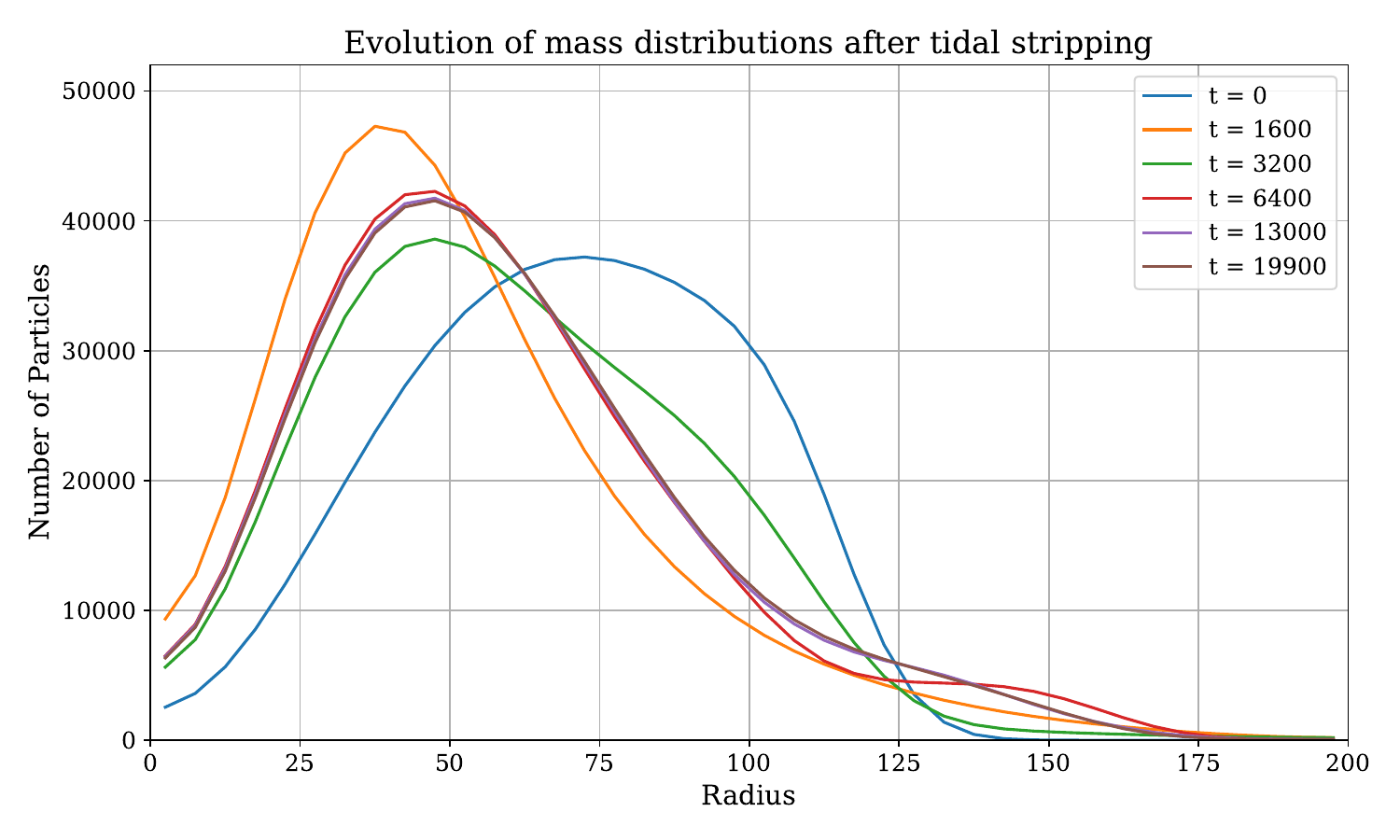}
     \caption{The time evolution of the mass distribution ($\propto r^2\, \rho(r)$ (in dimensionless units) of a self-consistent self-gravitating system that has had the outer 40\% of its mass stripped.  The system was evolved with 600,000 particles for 20 dynamical times with 1000 time steps per dynamical time.  The system relaxes to a density profile that is well approximated by $\rho(r) \propto (r^2+b^2)^{-5}$.}
\label{fig:tidalstripping}
\end{figure}

These questions are easily explored with this calculation.  To illustrate, we show the time evolution of the density profile of a dark-matter halo that at some particular time has the outer 40\% of the mass stripped away suddenly.  Results are shown in Fig.~\ref{fig:tidalstripping} for the evolution of a system of the 600,000 particles that remain in an initially self-consistent self-gravitating system of 1,000,000 after the outer 40\% are removed.  The evolution is shown for twenty dynamical times with 1000 time steps per dynamical time.  (This computation takes about 10 minutes on a 2023 MacBook Pro with an M2 pro chip.)  The density peak initially becomes higher and shifts to smaller radii, then to a bit larger radius and smaller amplitude, and then oscillates a few times before settling down, after $\sim10$ dynamical times to a density profile that is well described by $\rho(r) \propto (r^2+b^2)^{-5}$.  The fairly rapid approach to a new equilibrium is in accord with what is seen in full N-body simulations of halo mergers \cite{Wang:2020hpl}.

\begin{figure}[!htbp]
\includegraphics[width=\columnwidth]{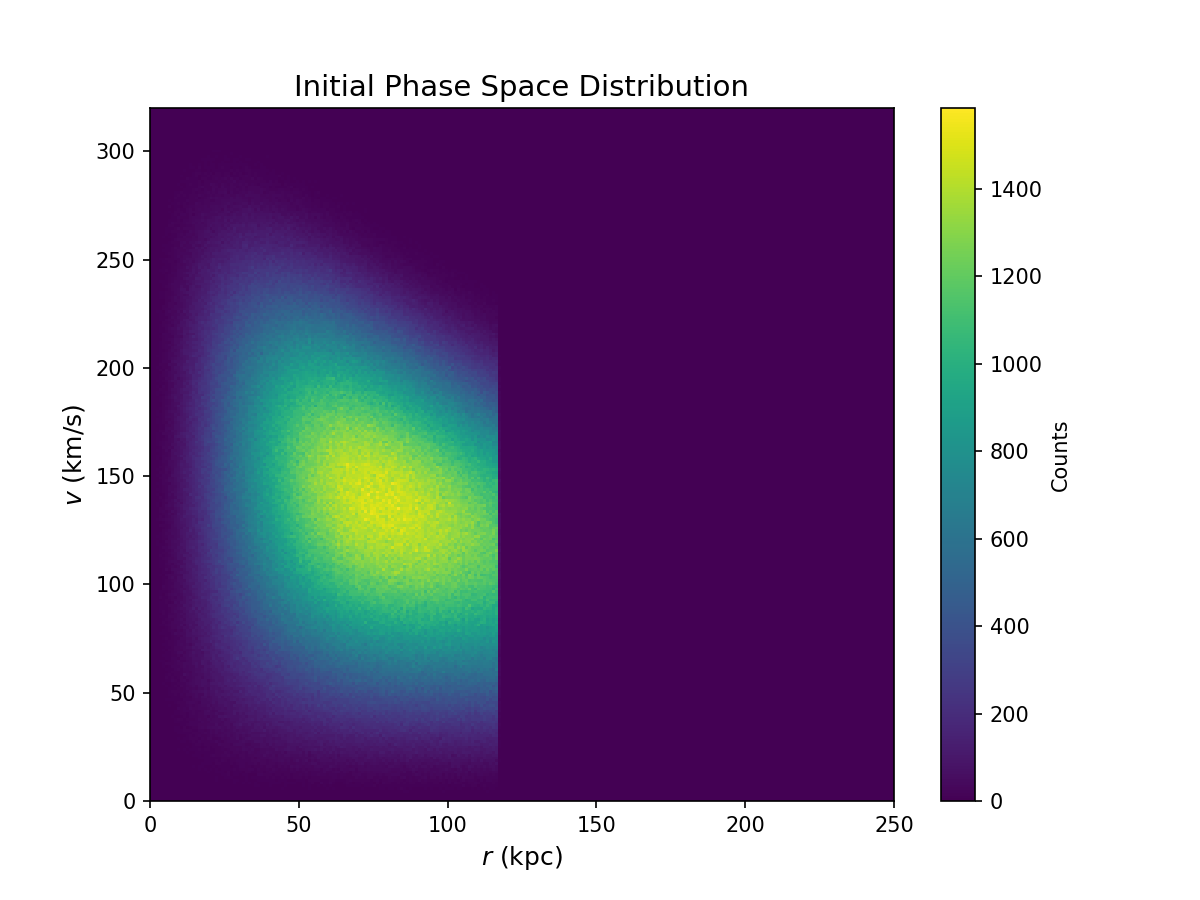}
\includegraphics[width=\columnwidth]{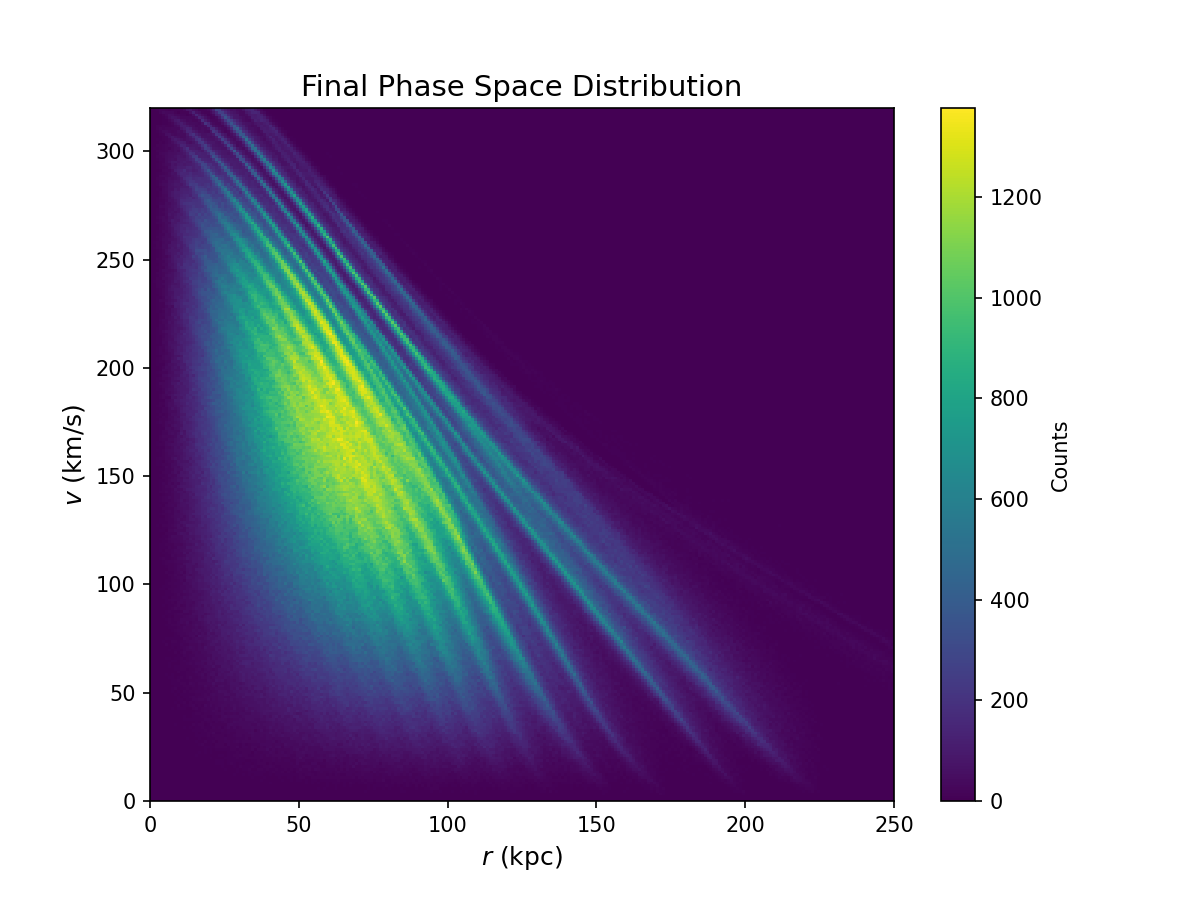}
     \caption{The phase-space distribution of the halo right after tidal stripping of the outer 40\% of particles (top) and after 25 dynamical times (bottom).}
\label{fig:phasespace}
\end{figure}

We also show in Fig.~\ref{fig:phasespace} a snapshot of the phase-space distribution (i.e., the number of particles in the $r$-$v$ plane) at the initial time (right after stripping) and after 25 dynamical times of evolution.  The resolution in this two-dimensional distribution was achieved with 12,000,000 particles, which takes something on the order of hours on a 2020 era computer, and it allows us to see winding in the phase space after tidal stripping.  An animation showing the evolution is provided on the GitHub.  We leave further investigation of the physics to future work.

\section{SMBH growth}
\label{sec:smbh}

\begin{figure}[!htbp]
\includegraphics[width=\columnwidth]{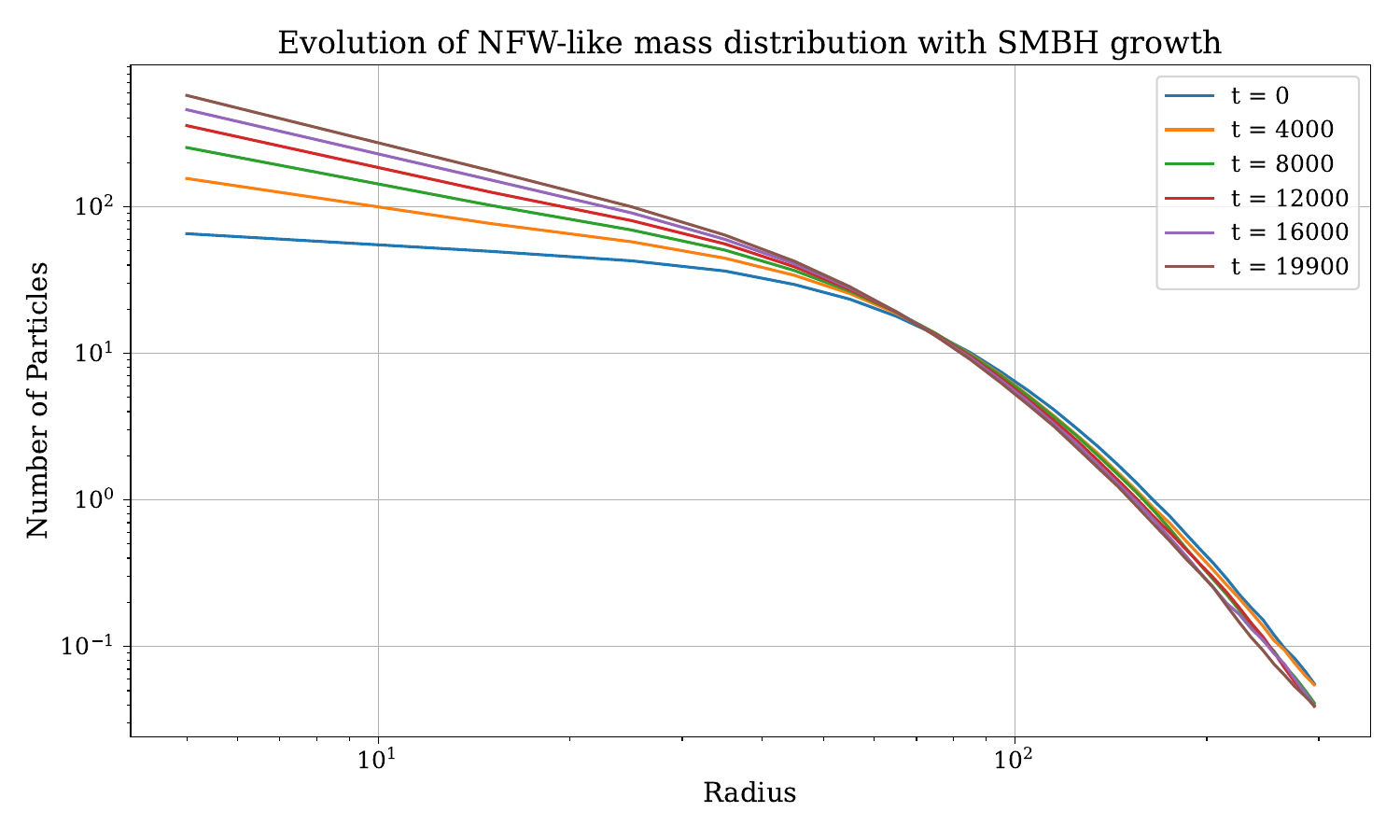}
     \caption{The evolution of the density distribution  (in dimensionless units) of a self-consistent self-gravitating system in which a supermassive black hole is slowly grown to 0.1 times the total halo mass over the integration time of 20 dynamical times.}
\label{fig:smbh}
\end{figure}

The effects of the growth of a supermassive black hole at the center of a collisionless system of particles have been considered in Refs.~\cite{Peebles:1972,Young:1980,Ipser:1987ru,Quinlan:1994ed,Gondolo:1999ef,Ullio:2001fb}.  Most of this work is advanced with a Fokker-Planck approach or with conservation laws and consistency conditions.  We show in Fig.~\ref{fig:smbh} the evolution of a halo in which a supermassive black hole is grown to 0.1 times the total halo mass over the integration time of 20 dynamical times.  The expected growth of a cusp near the center is seen here.  The matter distribution at large radii is also brought in.  The growth of the black hole introduces a far shorter orbital time scale for particles near the center than for particles further out, and so precise evolution of this cusp requires finer time steps.  It will be interesting to repeat this exercise to explore the growth of a $r^{-9/4}$ spike expected \cite{Gondolo:1999ef,Ullio:2001fb} in a halo that has initially an NFW \cite{Navarro:1996gj} profile.

\section{Conclusions}
\label{sec:conclusions}

We have presented a new computationally efficient approach to studying the N-body dynamics of spherically symmetric self-gravitating systems of collisionless particles that is easily augmented to study various departures from equilibrium and to include new-physics effects.  We have illustrated by numerically evolving halos with decaying dark matter, variable Newton's constant, the growth of a supermassive black hole, and the response of a self-gravitating system to stripping of its outer layers.  The latter numerical experiment finds a relaxation to a final state with a density profile that falls off as $r^{-10}$ at large radii $r$.  It will be interesting to explore more carefully in future work each of these examples.  The evolutions can be characterized more precisely and explored for a variety of different parameter combinations, different halo profiles, and also for velocity distributions that are not necessarily isotropic, as we have assumed here.  The implications for stellar orbits \cite{Sharma:2022qtw} of resulting mass rearrangements seen in time-evolving dark-matter halos can also be studied.  Here we have focused on density profiles, but there may also be interesting structure in the evolution of the velocity distributions.

The approach can also be used to explore the evolution of systems far from equilibrium and virialization/relaxation in gravitational collapse.  The approach is easily augmented \cite{Kamionkowski:2025} to study the effects of dark matter with other types of decays (e.g., two-body final states), models with elastic \cite{Spergel:1999mh,Dave:2000ar,Tulin:2017ara,Nishikawa:2019lsc} 
and inelastic \cite{Tucker-Smith:2001myb,Vogelsberger:2018bok} self interactions, or sticky dark matter \cite{Krnjaic:2014xza}.  The computational infrastructure for these self-interactions can also be used to follow the dynamical effects of annihilating dark matter.  It should also enable the inclusion of hard two-body gravitational scattering and thus (if we endow particles with masses) efficiently evolve dynamical mass segregation in globular clusters.

Certainly, there are many other applications that we have not listed, and it will be interesting to see the insights provided by future work that employs this new tool.

\begin{acknowledgments}
We acknowledge useful discussions with Heidi Newberg, Oren Slone, Scott Tremaine, Ben Wandelt, and Tomer Yavetz.  We use routines from the GNU Scientific Library \cite{GSL} and FFTW \cite{FFTW.jl-2005}. This work was supported at JHU by NSF Grant No.\ 2412361, NASA ATP Grant No.\ 80NSSC24K1226, the Guggenheim Foundation, and the Templeton Foundation. This work was supported at UBC by a Discovery Grant from NSERC of Canada. MK thanks the Center for Computational Astrophysics at the Flatiron Institute for hospitality. 
\end{acknowledgments}

\appendix

%%%%%%%%%%%%%%%%%%%%%%%%%%%%%%%%%%%%%%%%%%%%%%%%%%%%%%%%%%%%%%%%%%%%%%%%%%%%%
\section{Overall {\tt NSphere} Algorithm}
\label{app:nsphere_overview}

{\tt NSphere} models a spherical dark-matter halo as a collection
of concentric spherical shells of particles. Each shell is
characterized by its radius \(r_n\), its radial velocity \(v_n\),
and its angular momentum \(\ell_n\). In every time step the
algorithm proceeds in two stages:

\subsection{Sorting}
The particles are sorted in order of increasing \(r_n\) so that
the cumulative mass enclosed,
\begin{equation}
M(r)=\sum_{r_i\le r} m\,,
\end{equation}
can be computed efficiently simply by reading out the radial
rank of the shell at \(r_n\) (here each \(m\) is the mass of one
shell). By Newton's shell theorem the gravitational acceleration
acting on a particle depends only on \(M(r)\).

\subsection{State Update}
Once the particles are sorted, each particle is updated
independently. Thus, each particle’s update is equivalent to
solving a pseudo-Kepler problem with equations of motion
\begin{equation}
\dot{r}=v,\qquad \dot{v}=a(r)=-\frac{G\,M(r)}{r^2} +\frac{\ell^2}{r^3}\,
\end{equation}
where here and throughout the Appendix $v$ is the {\it radial} velocity, and $\ell=rv$ is the angular-momentum per unit mass.  In each time step the velocity and position are advanced using a velocity-based leapfrog integrator. A simple non-adaptive leapfrog update is given by
\begin{equation}
v\Bigl(t+\frac{\Delta t}{2}\Bigr)=v(t)+\frac{\Delta t}{2}\,a\bigl(r(t)\bigr),
\end{equation}
\begin{equation}
r(t+\Delta t)=r(t)+\Delta t\,v\Bigl(t+\frac{\Delta t}{2}\Bigr),
\end{equation}
\begin{equation}
v\bigl(t+\Delta t\bigr)=v\Bigl(t+\frac{\Delta t}{2}\Bigr)
+\frac{\Delta t}{2}\,a\bigl(r(t+\Delta t)\bigr)\,.
\end{equation}
(In the full algorithm the update is performed with an adaptive
refinement leapfrog (LF) integration scheme as described in
Appendix~\ref{sec:stepper}.)

The cumulative mass is computed by summing the mass \(m\) of all
particles with \(r_i\le r\); this is equivalent to reading out the
radial rank of the shell at \(r_n\).

A high-level pseudocode for the overall update is as follows:
\begin{lstlisting}
For each time step:
  1. Sort shell coordinates (r_1,...,r_N) by r.
  2. Compute M(r) from r rank.
  3. For each shell n:
         Update (v_n, r_n) via adapt. LF method
\end{lstlisting}

%%%%%%%%%%%%%%%%%%%%%%%%%%%%%%%%%%%%%%%%%%%%%%%%%%%%%%%%%%%%%%%%%%%%%%%%%%%%%
\section{Sorting Algorithms}
\label{sec:sort}

Accurate and efficient sorting of the particles by their radial coordinate is
critical for determining the enclosed mass. At every time step the
particle array is re-sorted in order of increasing \(r_n\) and the
cumulative mass \(M(r)\) is computed by reading out the radial rank of
each shell. Since the shells are nearly sorted at every step (because
of the locality of the time evolution), this operation is very efficient.

\subsection{QuadSort}
\verb|QuadSort| \cite{quadsort} is an efficient sorting algorithm chosen for our software because it quickly organizes nearly sorted data—a common case in our simulations where individual shells rarely cross in a single timestep. Instead of reordering the entire dataset, it first sorts small groups using a “quad swap” and then merges them, capitalizing on the existing order.

This adaptive strategy achieves nearly $O(N)$ performance for nearly sorted data while maintaining a worst-case of $O(N\log N)$, a huge improvement over naive $O(N^2)$ methods. Such efficiency is especially beneficial when handling very large particle numbers (typically $N > 10^6$), where sorting can dominate the computational cost\footnote{The precise turnover point at which sorting dominates depends on the specific computer architecture and specifications.}. Overall, its design minimizes extra processing and memory overhead, resulting in faster data handling and more responsive simulations.

\subsection{Insertion Sort}
Insertion sort is a straightforward algorithm that builds a sorted array one element at a time by inserting each new element into its proper place among those already sorted. In the worst-case scenario, insertion sort requires on the order of $O(N^2)$ comparisons and shifts. However, when the data is nearly sorted—as is typical in our simulations, where shells seldom cross between time steps—the algorithm's performance approaches $O(N)$.

For certain ranges of $N$ and on some hardware architectures, a parallelized version of insertion sort exhibits runtime comparable to that of \verb|QuadSort|, and in some cases even shorter. Its low overhead and simplicity make it particularly effective when only minor adjustments are needed between time steps. \footnote{For further details on the asymptotic behavior of insertion sort, see Refs.\ \cite{KnuthACP3,CormenAlgorithms}.}

\subsection{Parallelization}
Efficient sorting is critical for our simulations. Although parallelizing a sort is not generally advantageous for arbitrary datasets, our nearly sorted arrays do benefit from parallelization. We employ OpenMP (OMP) to run the sorting routines in parallel across all available CPU cores. The array of shell radii is divided into several segments that are sorted independently. At the boundaries between these segments, overlapping regions—chosen to be larger than the expected maximum number of shells a given shell can cross in a timestep \(\Delta t\)—are resorted to merge the independently sorted segments.

A high-level pseudocode for the parallel sorting is given below:
\begin{lstlisting}
Partition the array of shell radii into M segments.
#pragma omp parallel for
For each segment i = 1 to M:
    Sort segment i using QuadSort or Insertion Sort.

For each boundary between segments:
    Define an overlap region that extends beyond 
      the boundary by the maximum expected
      displacement in $\Delta$t.
    Sort the overlap region to merge the adjacent
      segments.
\end{lstlisting}

This parallelization strategy, leveraging OpenMP, reduces sorting time when handling very large particle numbers, thereby improving the overall efficiency of our simulations.

%%%%%%%%%%%%%%%%%%%%%%%%%%%%%%%%%%%%%%%%%%%%%%%%%%%%%%%%%%%%%%%%%%%%%%%%%%%%%
\section{Adaptive Refinement Leapfrog Integration}
\label{sec:stepper}

{\tt NSphere} advances particle states with a velocity-based leapfrog
integrator that employs an adaptive refinement leapfrog integration
scheme. In each full time step \(\Delta t\) the integrator subdivides
the interval into micro-steps and compares two independent estimates
of the new state.

\subsection{Adaptive Kick--Drift Ordering (Coarse and Fine)}
The update over a full time step \(\Delta t\) is carried out as
follows:
\begin{equation}
\begin{aligned}
\underbrace{\text{\scriptsize Kick}}_{\Delta t/(2N)\,a(r)}
&\longrightarrow
\underbrace{\text{\scriptsize Drift}}_{\Delta t/N\,v}
\longrightarrow
\underbrace{\text{\scriptsize Kick}}_{\Delta t/N\,a(r)}
\longrightarrow \cdots \\[1ex]
&\longrightarrow
\underbrace{\text{\scriptsize Drift}}_{\Delta t/N\,v}
\longrightarrow
\underbrace{\text{\scriptsize Kick}}_{\Delta t/(2N)\,a(r)}\,.
\end{aligned}
\end{equation}
A \textit{Kick} updates the velocity \(v\) via 
\(\Delta t/(2N)\,a(r)\) (or \(\Delta t/N\,a(r)\) in a full kick) and a
\textit{Drift} updates the position \(r\) via 
\(\Delta t/N\,v\). Note that in one complete leapfrog cycle the standard
update is a half-kick, followed by a full drift, and then a half-kick;
when this cycle is repeated, the two consecutive half-kicks (which occur
while \(r\) remains constant) combine to yield an effective full kick. This
produces the \(2N+1\) (coarse) pattern; doubling \(N\) (i.e., halving the
micro-step sizes) produces the \(4N+1\) fine estimate.

A high-level pseudocode for the adaptive refinement leapfrog
integration is:
\begin{lstlisting}
Set N = 1.
Repeat:
  (1) Coarse state (2N+1 steps):
      a. Half-kick:  v += $\Delta$t/(2N)*a(r)
      b. For i = 1 to N-1:
           Drift:   r += $\Delta$t/N*v
           Kick:    v += $\Delta$t/N*a(r)
      c. Final drift:  r += $\Delta$t/N*v
      d. Final half-kick: v += $\Delta$t/(2N)*a(r)
  (2) Fine state (4N+1 steps), as in (1).
  (3) If diff < tol, accept; else, set N = 2N.
Until converged.
\end{lstlisting}
The state at time \(t+\Delta t\) is given by the position
\(r(t+\Delta t)\) and the velocity \(v(t+\Delta t)\) after the
final half-kick.

\vspace{-0.4cm}
\subsection{Levi-Civita Regularization}
\vspace{-0.2cm}
For nearly radial (low-\(\ell\)) orbits the evolution near
pericenter can become stiff. To remedy this, {\tt NSphere} employs a partial
Levi-Civita coordinate transformation---historically introduced by
Levi-Civita \cite{LeviCivita1920} ---which is generally used to
remove or soften singularities in dynamical systems by mapping the original
coordinates into a regularized space. We first define a critical
radius
\begin{equation}
r_{\mathrm{crit}}=\frac{\alpha\,\ell^2}{G\,M(r)},
\end{equation}
with \(M(r)\) the mass enclosed within \(r\) and \(\alpha\) a tunable
parameter (we take \(\alpha=\tfrac{1}{20}\)). When a particle's radius
falls below \(r_{\mathrm{crit}}\) the integration is switched to a
Levi-Civita scheme.
After defining the new radial variable
\begin{equation}
\varrho=\sqrt{r}\,,
\end{equation}
and introducing a Sundman time reparameterization \cite{Sundman1912} 
\begin{equation}
\frac{dt}{d\tau}= r = \varrho^2\,,
\end{equation}
to a fictitious time \(\tau\) we evolve this transformed system.

While in Levi-Civita coordinates the mass enclosed $M(\varrho)$ is still defined by the global sorting in $r$ and the
evolution equations become
\begin{equation}
\frac{d\varrho}{d\tau}=\frac{1}{2}\,\varrho\,v,\qquad
\frac{dv}{d\tau}=-\frac{G\,M(\varrho)}{\varrho^2}+\frac{\ell^2}{\varrho^4}\,.
\end{equation}
Because each particle evolves independently after sorting---each
experiencing a pseudo-Kepler problem---a radial Levi-Civita
change of variables may be applied to each particle individually as
needed.\footnote{%
  Historically, Levi-Civita \cite{LeviCivita1920} introduced a conformal
  mapping $(x + i\,y) = u^2$, identifying $|u| = \sqrt{r} = \varrho$. In that
  full approach, one also sets $dt = |u|^2\,d\tau$, transforming the Kepler problem
  into an equivalent harmonic oscillator system and completely removing 
  the collision singularity. Here, we retain only the radial part 
  $\varrho = \sqrt{r}$ along with the Sundman reparameterization 
  \cite{Sundman1912} $dt = r\,d\tau = \varrho^2 d\tau$, 
  which is enough to soften the pericenter singularity. For more modern
  discussions of Levi-Civita regularization, see e.g., Refs.~\cite{Moser1969,Moser2001,Celletti2010,Waldvogel2010}.
}

Adaptive refinement leapfrog integration is then applied to the \(\varrho(\tau)\) coordinate and the
physical velocity \(v(\tau)\) with the physical time
\(t(\tau)\) updated alongside. In particular, the physical
time step \(\Delta t\) is mapped to several corresponding fictitious
time steps \(\Delta\tau\) that span the interval; if the accumulated
time overshoots \(t+\Delta t\), linear interpolation is used to compute
a fractional final step so that the evolution proceeds exactly from
\(t\) to \(t+\Delta t\).

\vspace{-0.0cm}
\bibliography{refs}% Produces the bibliography via BibTeX.

\end{document}